\documentclass[a4paper,11pt]{article}
\usepackage{pos,slashed,dsfont}

\newcommand{\Z}{\mathcal{Z}}
\newcommand{\dd}{\textmd{d}}

\title{Thermal QCD with external imaginary electric fields on the lattice}

\author*[a]{Gergely Endr\H{o}di}
\author[a]{Gergely Mark\'o}

\affiliation[a]{Fakult\"at f\"ur Physik, Universit\"at Bielefeld\\
  D-33615 Bielefeld, Germany.}

\emailAdd{endrodi@physik.uni-bielefeld.de}
\emailAdd{gmarko@physik.uni-bielefeld.de}

\abstract{We study QCD at finite temperature in the presence of imaginary electric fields. In particular, we determine the electric susceptibility, the leading coefficient in the expansion of the QCD pressure in the imaginary field. Unlike for magnetic fields, at nonzero temperature this coefficient requires a non-trivial separation of genuine electric field-related effects and spurious effects related to the chemical potential, which becomes an unphysical gauge parameter in this setting. Our results are based on lattice simulations with stout improved dynamical staggered quarks at physical quark masses.}

\FullConference{%
 The 38th International Symposium on Lattice Field Theory, LATTICE2021
  26th-30th July, 2021
  Zoom/Gather@Massachusetts Institute of Technology
}

\begin{document}
\maketitle

\section{Introduction}

It is by now well established that the early stages of off-central heavy-ion collisions exhibit strong electromagnetic fields~\cite{Kharzeev:2012ph}, even if only for a short period of time~\cite{Tuchin:2013apa,McLerran:2013hla}. 
The magnetic components received most attention 
due to their possible phenomenological impact via, e.g., the chiral magnetic effect~\cite{Fukushima:2008xe}.
On an event-by-event basis, the electric components can however be as large as the magnetic 
ones, reaching up to values $\propto \mathcal{O}(m_\pi^2)$~\cite{Voloshin:2010ut,Deng:2014uja} and therefore affecting 
strong interaction processes in a potentially significant way. Electric 
fields are expected to be particularly relevant for asymmetric systems (e.g.\ copper on gold at RHIC)~\cite{Voronyuk:2014rna}.

In this talk we do not attempt to describe the highly complex out-of-equilibrium system
including time-dependent electromagnetic fields. Our focus lies in the treatment of 
equilibrium strongly interacting matter in the presence of static, homogeneous background
fields at nonzero temperature on the lattice.
For background {\it magnetic} fields $B$ the most important aspects, like the phase diagram~\cite{DElia:2010abb,Bali:2011qj,Endrodi:2015oba} and the equation of state~\cite{Bali:2014kia,Bali:2020bcn} have already been investigated in detail. For external {\it electric} fields $E$ our knowledge is mostly limited to low-temperature effects 
like electric polarizabilities of hadrons, see, e.g., Refs.~\cite{Engelhardt:2007ub,Lujan:2014kia}.
One lattice study discussed charge separation in an unphysical system with isospin electric charges for 
quarks both at low and at high temperature~\cite{Yamamoto:2012bd}. 
Further thermodynamic aspects have been studied
perturbatively at high temperature, see e.g.~\cite{Gies:1998vt}.

In comparison to the magnetic case, lattice simulations involving external electric fields entail several new conceptual challenges. First, a constant electric field accelerates charged particles, bringing the system inevitably out of equilibrium, naively inaccessible for standard simulations. Second, as we will see below, at nonzero temperature
electric fields automatically enforce an averaging over the chemical potential $\mu$, implying a mixing between $E$ and $\mu$.
Third, the QCD action becomes complex in the presence of a real electric field, hindering importance sampling-based simulations.
Here we aim to solve the first two of these issues and develop 
an approach that enables an equilibrium discussion and also eliminates the mixing between $E$ and $\mu$. We also discuss a Taylor-expansion method in imaginary electric fields in order to address the complex action problem, but we will not perform 
an analytic continuation to real electric fields here.

\section{Equilibrium and mixing with the chemical potential}
\label{sec:equ}

\begin{figure}[t]
 \centering
 \includegraphics[width=8cm]{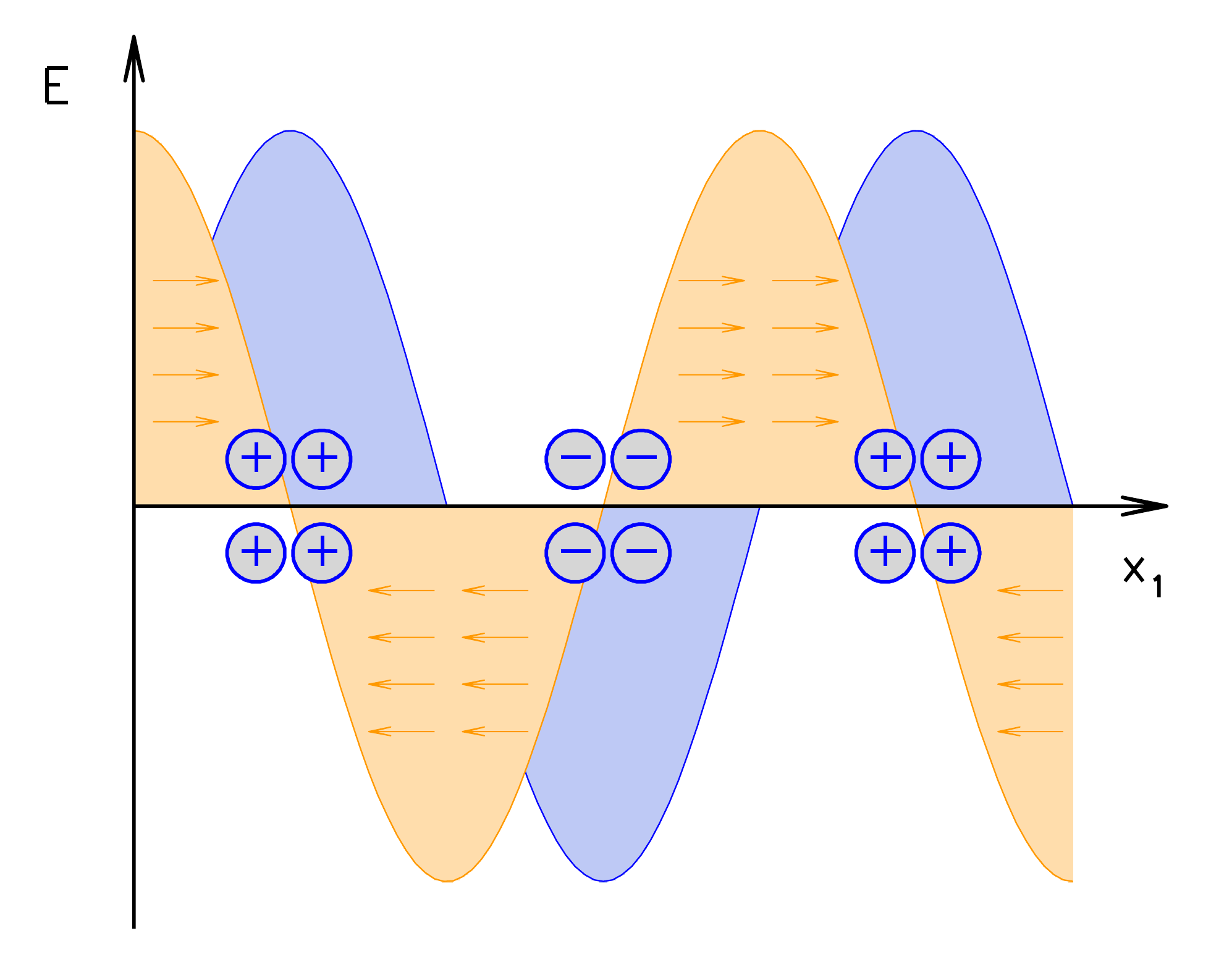}
 \caption{\label{fig:demo1}Illustration of the equilibrium situation in the presence of an 
 electric field $E \parallel x_1$ modulated in the $x_1$ direction 
 (dark yellow area and arrows). 
 Positive (negative) electric charges accumulate at points with $E=0$ pointed 
 towards (away from) the electric field, giving rise to the charge density profile 
 indicated by the blue region.
}
\end{figure}

Instead of a homogeneous electric field, let us consider an oscillatory field of the form 
$\mathbf{E}(x_1)=E \cos(p_1 x_1)\,\mathbf{e}_1$, pointing in the $x_1$ direction and modulated with a wavelength $1/p_1$, see Fig.~\ref{fig:demo1}. In this setting, 
the electrically charged particles 
will move around until the effect of the electric field is balanced by the interactions between them -- i.e.\ the strong force as well as the electromagnetic repulsion (although in this contribution we do not take dynamical QED effects into account).
Thus, an equilibrium will be reached and lattice simulations of this 
setup will correspond to this equilibrium and fluctuations around it. 
The resulting modulated distribution of the electric charge density (see also Fig.~\ref{fig:demo1}) arises 
due to electric polarization and is described by the electric permittivity of the QCD medium.
Once we have approached the infinite volume limit, we can extrapolate the momentum $p_1$ to zero. For QCD in the absence of dynamical QED effects, interactions fall off exponentially with the distance and the $p_1\to0$ limit is expected to be smooth already on sufficiently large volumes -- we get back to this point below. With this approach we can therefore 
investigate the impact of constant electric fields via a smooth limit of 
equilibrium lattice simulations.

To quantify the electric polarization effect, we need to compare the equilibria at $E=0$ and at a small but nonzero $E\neq0$.
This leads us to the second issue -- the mixing of the effect of the 
electric field and that of the chemical potential at nonzero temperature $T$.
Here we consider a constant electric field (referring to the above discussion, one may think of a modulated field with long wavelength and zooming into the vicinity of the origin). 
To avoid time-dependent electromagnetic potentials, we consider the static gauge $A_0=E x_1$ for the electric field.
Now on the one hand, at $E=0$ the free energy density $f=-T/V\cdot \log\Z$ depends on $\mu$. 
On the other hand, for $E\neq0$ the chemical potential becomes a mere gauge degree
of freedom, since the potentials $A_0=E x_1$ and $A_0=E x_1+\mu$ only differ 
by an electromagnetic gauge transformation. We are lead to conclude that $f$, being 
a gauge invariant quantity, cannot depend on $\mu$. In other words, at $E\neq0$ the chemical potential is averaged over automatically. To meaningfully compare 
the free energies at $E\neq0$ and at $E=0$, we therefore have to average over 
the chemical potentials also for the latter case.

The same effect also arises 
for imaginary (or, Euclidean) electric fields $\hat E$, considered
in the gauge $A_4=\hat E x_1$, and imaginary chemical potentials $\hat\mu$ (from here on we will only work with Euclidean parameters, denoted 
by the hat).
We demonstrate this by calculating the free energy density on the lattice 
for imaginary electric fields. The latter enter the Dirac operator for a quark with flavor $q$
via $\mathrm{U}(1)$ links $u_\nu$ in the $x_1-x_4$ plane, of the form~\cite{Detmold:2009dx}
\begin{equation}
u_4(n) = e^{i a^2 q \hat E n_1}, \qquad
\left.u_1(n)\right|_{n_1=N_1-1} = e^{-i a^2 q \hat E N_1 n_4}, \qquad
u_\nu(n) = 1 \quad\textmd{otherwise},
\end{equation}
where $a$ is the lattice spacing and $n_\nu=0\ldots N_\nu-1$ label the sites of the lattice. For periodic boundary conditions, the electric field 
is quantized as~\cite{tHooft:1979rtg}
\begin{equation}
a^2q\hat E  = \frac{2\pi N_E}{N_1 N_4}, \qquad N_E\in\mathds{Z}\,.
\label{eq:quantization}
\end{equation}
The linear size and the temperature are given by $L=N_1a$ and $T=(N_4a)^{-1}$.
The quark charges $q_f$ will be measured in units of the elementary electric charge $e$.

\begin{figure}[t]
 \centering
 \mbox{
 \includegraphics[height=5.8cm]{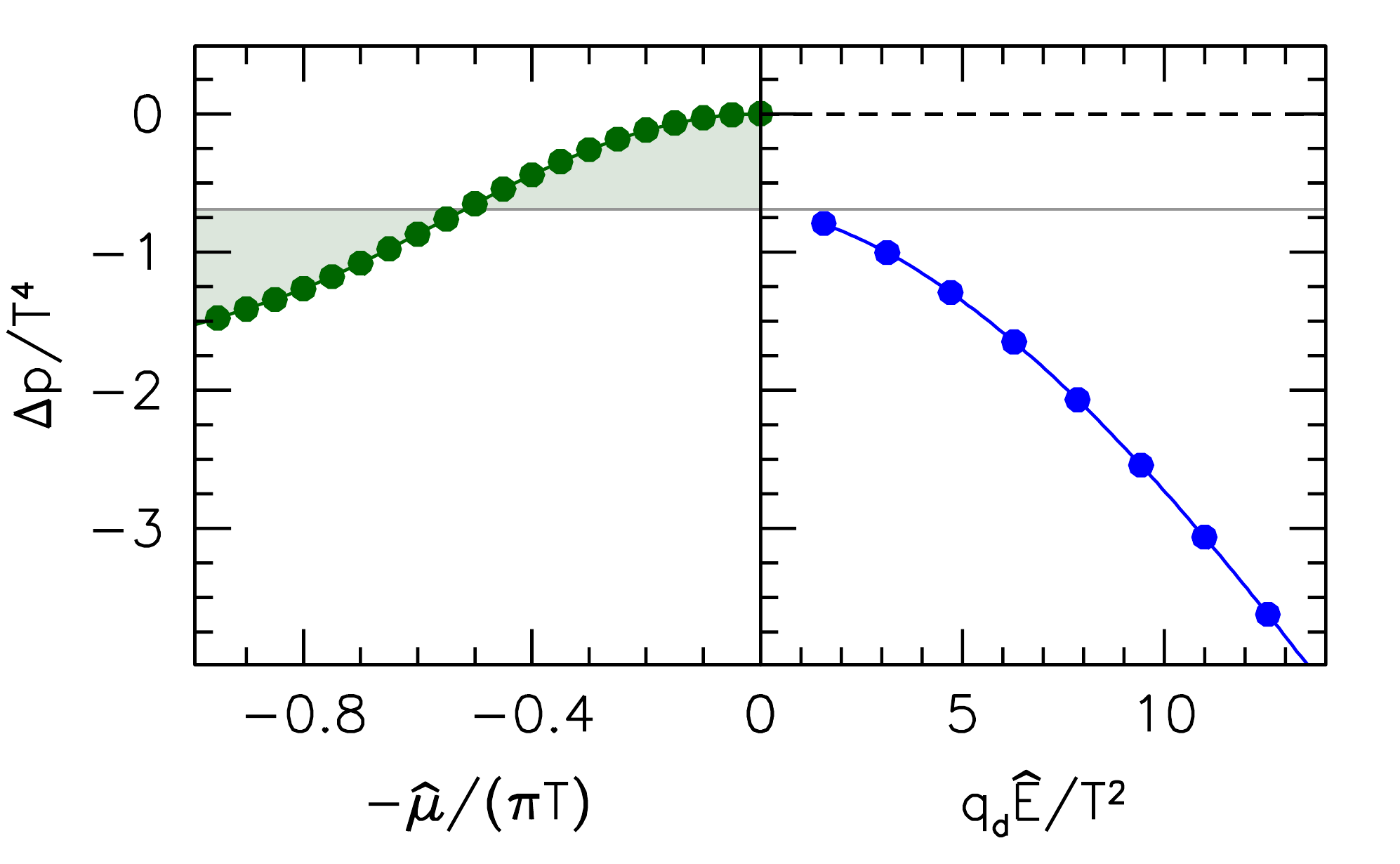}  
 \includegraphics[height=5.8cm,trim=210 0 0 0,clip]{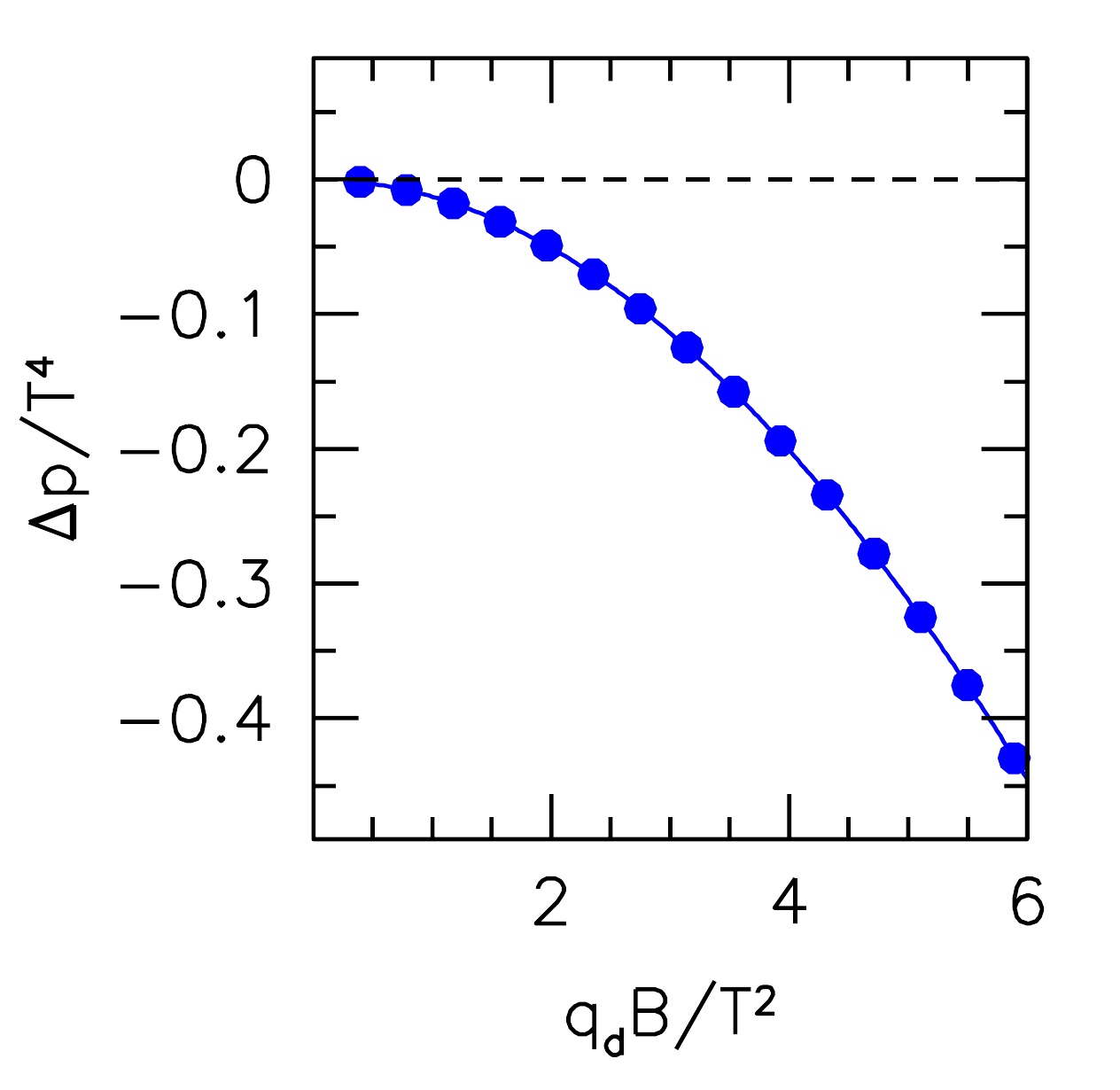} 
 }
 \caption{\label{fig:mismatch}The change of the pressure in units of $T^4$ due to the imaginary chemical potential (left panel), the imaginary electric field (middle panel) and the magnetic field (right panel) in the free case at high temperature. The measurements at quantized fields (blue dots) are connected merely to guide the eye. The measurements at $i\mu\neq0$ (green dots) are averaged over all chemical potentials (gray line).}
\end{figure}

For the moment we neglect gluonic interactions so that we may calculate
the free energy density by 
direct diagonalization of the Dirac operator for a colorless fermion. We employ the staggered 
discretization on a $24^3\times6$ lattice with $m/T=0.08$. The results for 
the change of the pressure, $\Delta p=-f + f(\hat E=0,\hat\mu=0)$, are shown in the middle 
panel of Fig.~\ref{fig:mismatch}. Notice the mismatch between the direct evaluation at $\hat E=0$ (i.e.\ $\Delta p=0$ there) and the continuation $\hat E\to0$ using the data at $\hat E>0$ (pointing clearly to $\Delta p<0$). As we have seen above, the correct $\hat E=0$ value is the one obtained 
by averaging over all imaginary chemical potentials $0<\hat\mu<\pi T$, shown in the left panel of the figure.\footnote{Note that here we discuss the free (colorless) case so no Roberge-Weiss-type transitions are present. The $2\pi T$ periodicity and the evenness of $\Delta p$ in $\hat\mu$ imply that the relevant interval is $0<\hat\mu<\pi T$, plotted in the figure. Also note that averaging over $\hat\mu$ amounts to considering the expansion of the grand canonical $\Delta p(\hat\mu)$ in the fugacity variable $e^{i\hat\mu N/T}$ and projecting to the $N=0$ sector.} The average of $\Delta p$ 
can now be compared meaningfully to
the $\hat E>0$ results.
Note that for magnetic fields no such mismatch is observed and the $B\to0$ results 
extrapolate back to $\Delta p=0$.

The possible values of the imaginary electric field become dense in the thermodynamic limit; in particular the minimal electric field $q\hat E_{\rm min}=2\pi T/L$ approaches zero. 
This implies that for the susceptibility with respect to the imaginary electric field,
to be defined below in~\eqref{eq:elsusc},
the mismatch term gives rise to an infrared divergence $\propto T^4/\hat E_{\rm min}^2\propto (LT)^2$.

\section{Taylor expansion}

In the calculations that follow a central role will be played by the electromagnetic
current-current correlator (written down before the path integral for quarks is carried out),
\begin{equation}
 G_{\mu\nu}(x_1)=\int \!\dd x_2 \,\dd x_3 \,\dd x_4 \,\langle j_\mu(x) j_\nu(0) \rangle\,, \qquad j_\mu=\bar\psi\gamma_\mu\psi\,.
\label{eq:Gmunu}
\end{equation}
Since the imaginary chemical potential multiplies $i$ times the integral of $j_4$ in the action, the 
(dimensionless) quark number susceptibility can be calculated in terms of the correlator as
\begin{equation}
c_2 \equiv \frac{1}{T^2}\cdot\frac{T}{V} \left.\frac{\partial^2 \log\Z}{\partial \hat\mu^2}\right|_{\hat\mu=0} = - \frac{1}{TV}\int \!\dd^4 x \,\dd^4 y \,\langle j_4(x) j_4(y) \rangle = -\frac{1}{T^2}\int \!\dd x_1 \,G_{44}(x_1)\,,
\label{eq:c2def}
\end{equation}
where we exploited the translational invariance of the expectation value. (Note that $c_2<0$ for this definition.)

For our periodic boundary conditions, we cannot differentiate with respect to $\hat E$ due to the quantization~\eqref{eq:quantization}. Let us for the moment switch to open boundary conditions in the $x_1$ direction (and choose our coordinate system so that $-L/2\le x_1 < L/2$). Then we are allowed to repeat the above derivation. The imaginary electric field multiplies
$ie$ times the integral of $j_4\cdot x_1$ in the action, so that the susceptibility in this setup (denoted by the bar) reads
\begin{equation}
 \bar\xi \equiv \frac{T}{V}\left.\frac{\partial^2 \log\Z}{\partial (e\hat E)^2}\right|_{\hat E=0}
 = -\frac{T}{V}\int \!\dd^4 x \,\dd^4 y \,x_1 y_1 \,\langle j_4(x)j_4(y)\rangle
  = -\frac{1}{L}\int \!\dd x_1 \,\dd y_1 \,x_1 y_1 \,G_{44}(x_1-y_1)\,.
\label{eq:elsusc}
\end{equation}
Replacing $x_1y_1=(x_1^2+y_1^2)/2-(x_1-y_1)^2/2$, exploiting again the translational (and this time also parity) invariance of the correlator and denoting $z_1=x_1-y_1$, 
we arrive at
\begin{equation}
 \bar\xi
 = -\frac{1}{L} \int \dd x_1 \,x_1^2\int \!\dd z_1 \,G_{44}(z_1) 
+ \frac{1}{L} \int \!\dd x_1 \int \! \dd z_1 \, \frac{z_1^2}{2} \,G_{44}(z_1)
= \frac{c_2}{12} (LT)^2 + \int \! \dd z_1 \, \frac{z_1^2}{2} \,G_{44}(z_1)\,,
\label{eq:naive}
\end{equation}
where we recognize the infrared divergent mismatch term $\propto (LT)^2$ that we anticipated above.

How to extract the physical part of $\bar\xi$? To answer this question we again need to 
consider oscillatory field profiles. For these, the Taylor-expansion in the amplitude 
of the field -- being a continuous variable -- is well-defined even for periodic volumes. Moreover, we can extrapolate to constant fields after 
the thermodynamic limit has been taken. For this, large volumes are already 
expected to be sufficient, see the discussion
regarding equilibrium in Sec.~\ref{sec:equ}. Recently we applied this 
method for magnetic fields in Ref.~\cite{Bali:2020bcn}. 
For the magnetic susceptibility, in 
the $A_2=B x_1$ gauge, the result reads
\begin{equation}
 \chi \equiv \frac{T}{V} \left.\frac{\partial^2 \log\Z}{\partial (eB)^2}\right|_{B=0} 
 = \int_0^{L/2}\! \dd z_1 z_1^2 \,G_{22}(z_1)\,.
\label{eq:chiobs}
\end{equation}
The observable~\eqref{eq:chiobs} may be recognized as the scalar vacuum polarization
that enters the calculation of the muon anomalous magnetic moment~\cite{Bali:2015msa} -- only 
here we need to calculate it at nonzero temperature. (For alternative methods to determine $\chi$, see for example~\cite{Bali:2013txa,Bali:2014kia}.)
The calculation for imaginary electric fields is completely
analogous, except that this time $A_4=\hat E x_1$ and therefore the density-density correlator appears,
\begin{equation}
 \xi = \int_0^{L/2}\! \dd z_1 z_1^2 \,G_{44}(z_1)=\bar\xi-\frac{c_2}{12}(LT)^2\,,
 \label{eq:xiphys}
\end{equation}
which is just the second term in our naive calculation in~\eqref{eq:naive}.
This is the physical contribution to the electric susceptibility. Note that the correlators $G_{\mu\nu}$ fall off exponentially with the distance, which suppresses finite volume effects in~\eqref{eq:chiobs} and~\eqref{eq:xiphys}, see~\cite{Bali:2020bcn} for more details.

\section{Results}

\begin{figure}[t]
 \centering
 \includegraphics[width=8cm]{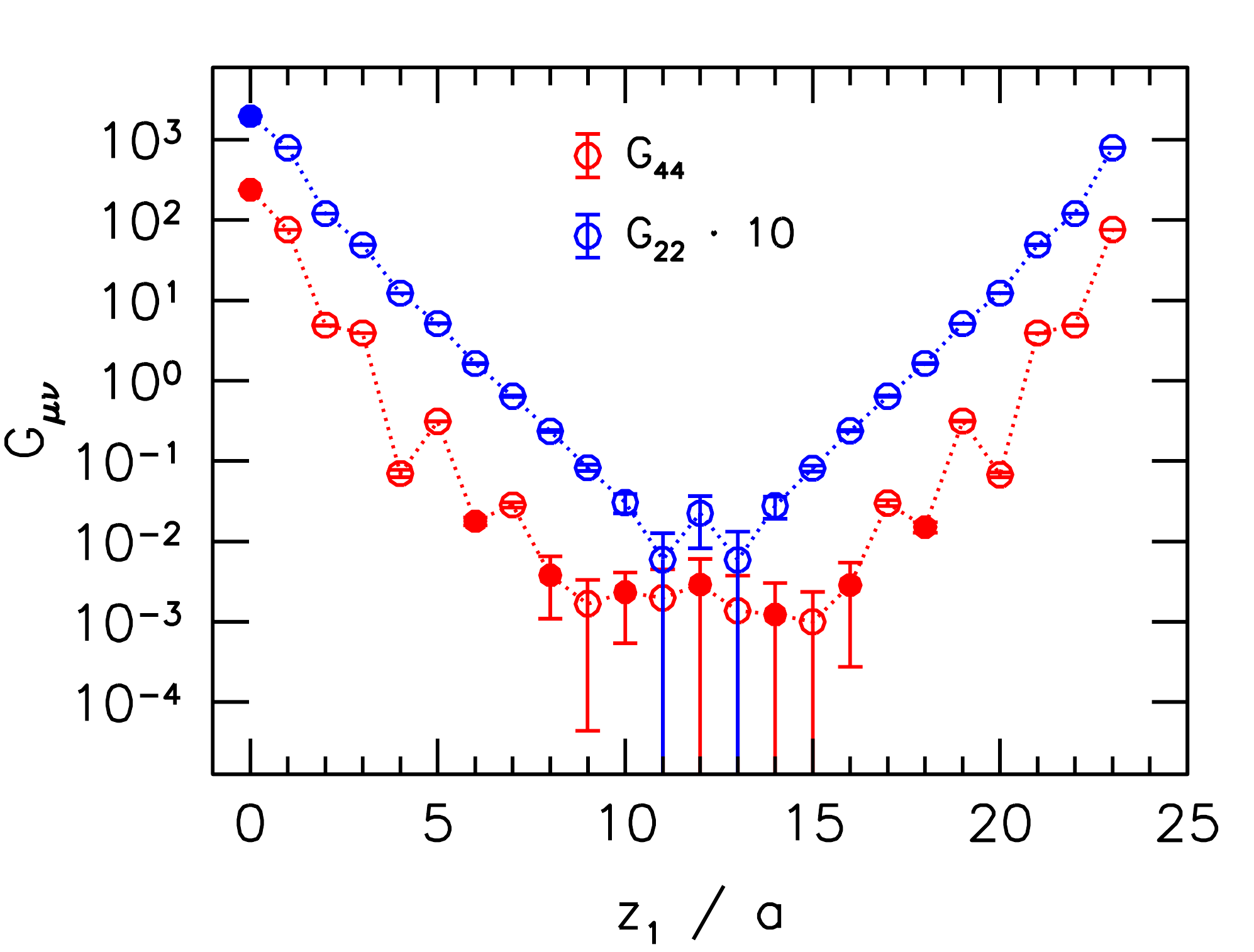} 
 \caption{\label{fig:corr}Current-current $G_{22}$ (blue) and density-density $G_{44}$
 (red) correlators at $T\approx 176 \textmd{ MeV}$ on our $N_t=6$ lattices. The former has been multiplied by a factor of $10$ for better visibility. 
 Filled (open) points indicate positive (negative) values. }
\end{figure}

We use the tree-level improved Symanzik gauge action and three flavors of 
stout smeared staggered quarks with physical masses. The current-current and density-density correlators~\eqref{eq:Gmunu} were determined
on $24^3\times6$ lattices for a range of temperatures. We calculated both 
the connected and the disconnected contributions to $G_{\mu\nu}$. To obtain precise 
results we used up to a thousand noisy estimators, localized to three-dimensional
$x_1$-slices. For the details of the measured operators, see Ref.~\cite{Bali:2015msa}. In Fig.~\ref{fig:corr} 
we plot the correlators at $T\approx176\textmd{ MeV}$, where the difference 
between the spatial and temporal components is already considerable.

\begin{figure}[t]
 \centering
 \mbox{
 \includegraphics[width=7.5cm]{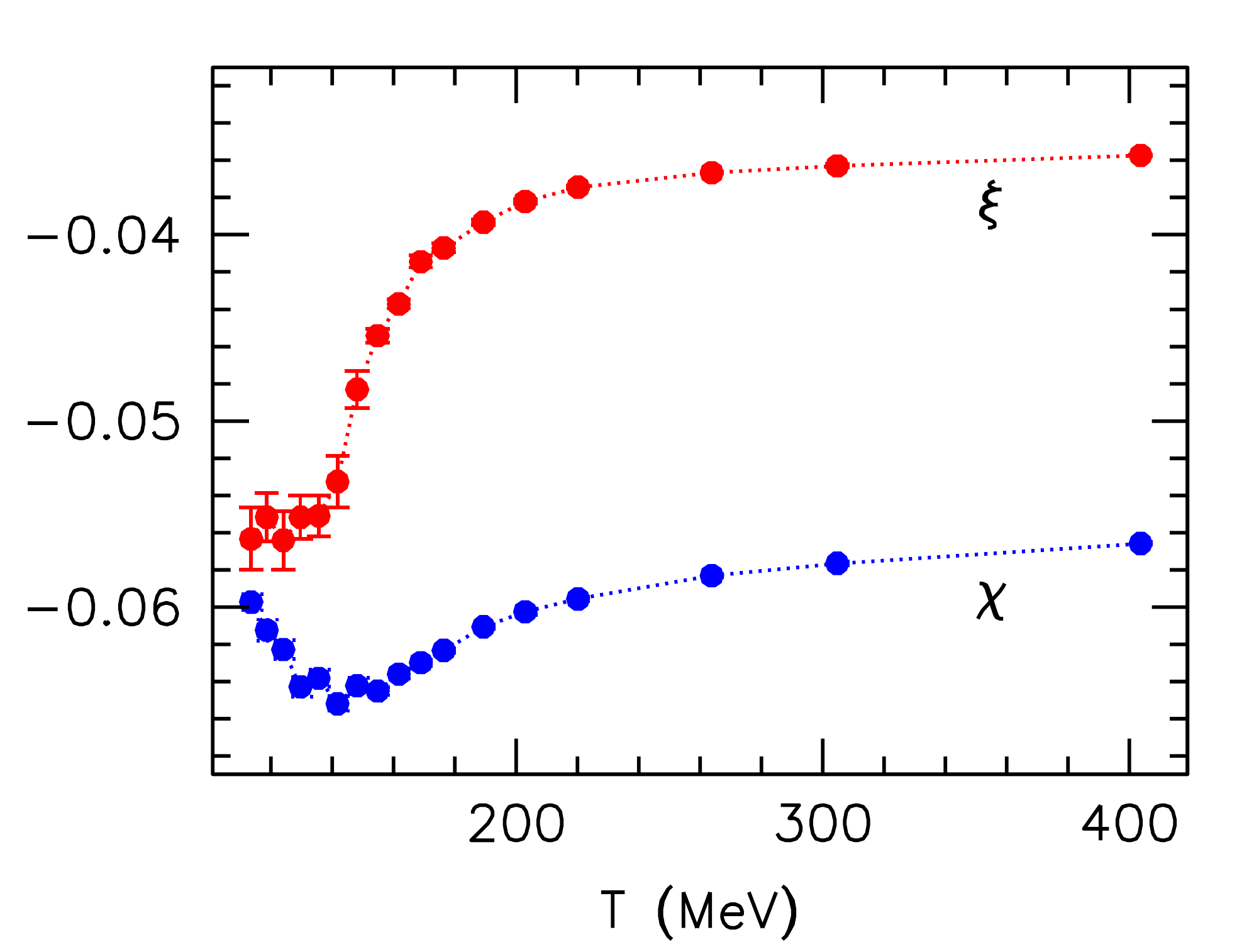} \;
 \includegraphics[width=7.5cm]{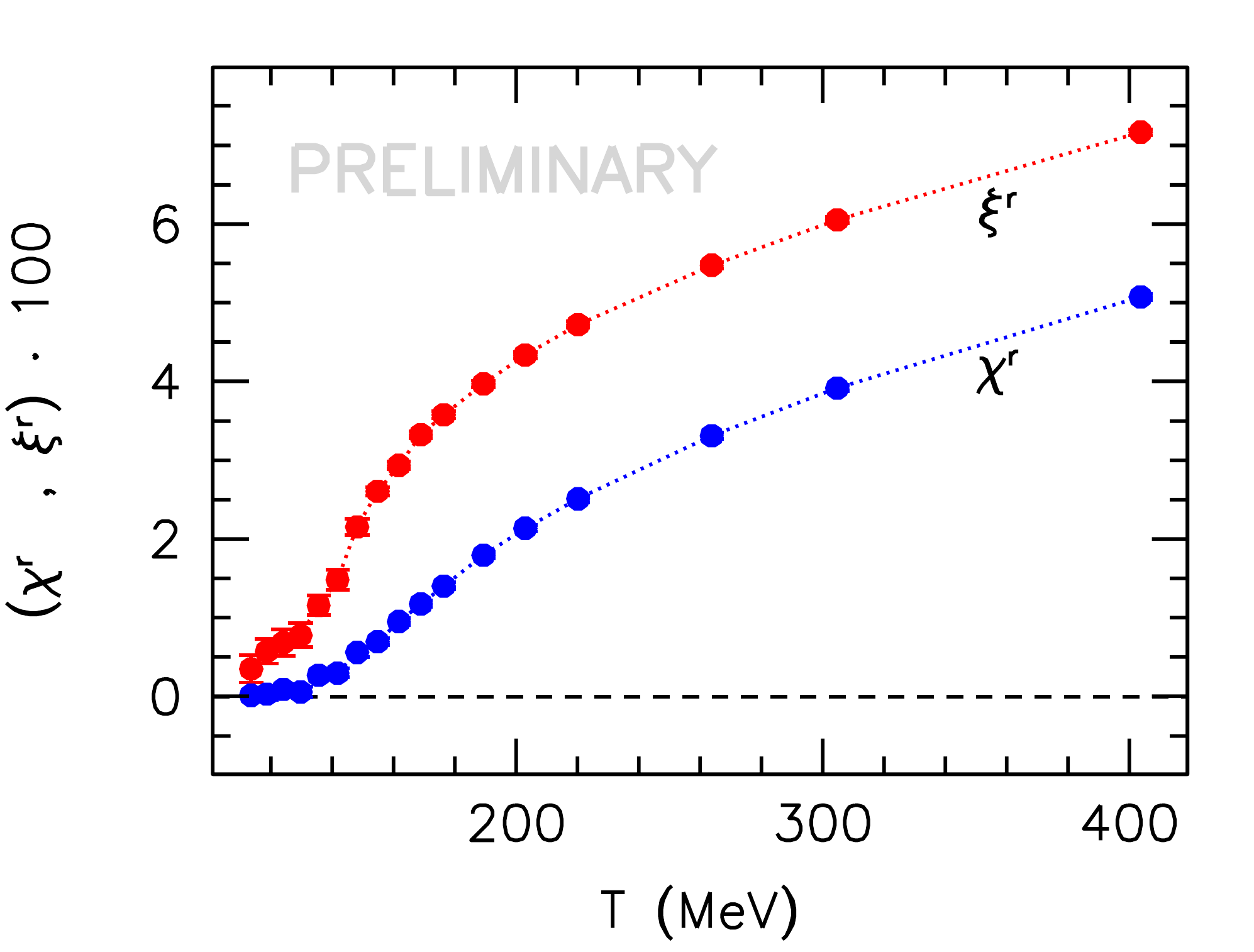} 
 }
 \caption{\label{fig:susc}Bare (left panel) and renormalized (right panel) 
 susceptibilities as a function of the temperature on $24^3\times 6$ lattices.
 Susceptibilities with respect to magnetic (blue) and imaginary electric (red)
 fields are both shown.}
\end{figure}

To calculate the susceptibilities~\eqref{eq:chiobs} and~\eqref{eq:xiphys} the 
correlators are convoluted with the quadratic kernel. The results are plotted 
in the left panel of Fig.~\ref{fig:susc} for several temperatures. These observables 
contain an additive divergence in the lattice cutoff (similar to the scalar 
vacuum polarization that enters the determination of the muon anomalous magnetic 
moment). To perform the additive renormalization we subtract the value of the 
bare susceptibilities evaluated at the same lattice spacing but on a $T\approx0$ ensemble (note that at $T=0$ the magnetic and electric susceptibilities are equal 
due to Lorentz symmetry). For this we employ the results of Refs.~\cite{Bali:2014kia,Bali:2020bcn}. 
This defines the renormalized susceptibilities
\begin{equation}
\chi^r=\chi(T)-\chi(T=0), \qquad \xi^r=\xi(T)-\xi(T=0) \,,
\end{equation}
which are plotted in the right panel of Fig.~\ref{fig:susc}. The renormalized magnetic susceptibility has been determined in this manner for a set of lattice spacings and a continuum extrapolation was performed in Ref.~\cite{Bali:2020bcn}.
This revealed negative values for low temperatures, indicative of a diamagnetic
response (due to pions). On the other hand, at high temperatures $\chi^r$ is positive, revealing paramagnetism (due to quarks). In the electric case
we have yet to perform the analytic continuation to obtain the susceptibility 
with respect to real electric fields.

\section{Summary}

In this contribution we discussed the response of the thermal QCD
medium to external electric fields, quantified in terms of the electric 
susceptibility. We demonstrated how to define the effect 
of the electric field by considering solely equilibrium simulations involving
oscillatory field profiles. We also pointed out that at nonzero temperatures 
the electric field mixes with the chemical potential and contributes an infrared 
divergent term to the susceptibility. This divergence can be avoided if 
an averaging over the chemical potential variable at zero electric field 
is carried out.
Finally we developed a Taylor-expansion scheme in the imaginary
electric field, similarly to the recent approach for magnetic fields~\cite{Bali:2020bcn} and presented first results for the susceptibility.\\

\noindent
{\bf Acknowledgments } 
This research was funded by the DFG (Emmy Noether Programme EN 1064/2-1 and the
the Collaborative Research Center CRC-TR 211 ``Strong-interaction
matter under extreme conditions'' -- project number
315477589 - TRR 211).

\bibliographystyle{utphys}
\bibliography{electric_pos}

\end{document}